\documentclass[twocolumn,showpacs,amsmath,amssymb]{revtex4}
\usepackage{graphicx}
\newcommand{\ltsim}{\protect\raisebox{-0.5ex}{$\:\stackrel{\textstyle <}{\sim}\:$}}
\newcommand{\gtsim}{\protect\raisebox{-0.5ex}{$\:\stackrel{\textstyle >}{\sim}\:$}}
\begin{document}

\title{Entanglement Entropy and Entanglement Spectrum for Two-Dimensional Classical Spin Configuration}
\author{Hiroaki Matsueda}
\affiliation{Sendai National College of Technology, Sendai 989-3128, Japan}
\date{\today}
\begin{abstract}
In quantum spin chains at criticality, two types of scaling for the entanglement entropy exist: one comes from conformal field theory (CFT), and the other is for entanglement support of matrix product state (MPS) approximation. They indicates that the matrix dimension of the MPS represents a length scale of spin correlation. On the other hand, the quantum spin-chain models can be mapped onto two-dimensional (2D) classical ones. Motivated by the scaling and the mapping, we introduce new entanglement entropy for 2D classical spin configuration as well as entanglement spectrum, and examine their basic properties in Ising and $3$-state Potts models on the square lattice. They are defined by the singular values of the reduced density matrix for a Monte Carlo snapshot. We find scaling relations concerned with length scales in the snapshot at $T_{c}$. There, the spin configuration is fractal, and various sizes of ordered clusters coexist. Then, the singular values automatically decompose the original snapshot into a set of images with different length scale. This is the origin of the scaling. In contrast to the MPS scaling, long-range spin correlation can be described by only few singular values. Furthermore, we find multiple gaps in the entanglement spectrum, and in contrast to standard topological phases, the low-lying entanglement levels below the gap represent spontaneous symmetry breaking. Based on these observations, we discuss about the amount of information contained in one snapshot in a viewpoint of the CFT scaling.
\end{abstract}
\maketitle

\section{Introduction}

The entanglement entropy is a common language connecting among various fields such as quantum information, quantum gravity, and condensed matter physics. The main reason of this wide applicability comes from a fact that the entropy picks up universality irrespective of details of their models. The entropy represents the amount of information across the boundary between a subsystem $A$ of linear size $L$ and its environment $B$. Starting with a wave function of the total system $\left|\psi\right>$, we first define the density matrix of $A$ by $\rho_{A}=tr_{B}\left|\psi\right>\left<\psi\right|$, which traces out degree of freedom inside of $B$. Then, the entanglement entropy $S_{A}$ is given by
\begin{eqnarray}
S_{A}=-tr_{A}(\rho_{A}\log\rho_{A}). \label{EE}
\end{eqnarray}
It has been extensively examined how this entropy behaves as functions of $L$ and spatial dimension $d$.

A well-known formula is called 'area-law scaling', $S\propto L^{d-1}$, which tells us non-extensivity of $S$ in contrast to the thermal entropy. This formula was originally introduced in a context of black-hole physics (Bekenstein-Hawking entropy)~\cite{Bekenstein,Hawking,Srednicki}, and examination of the scaling and its violation has been a hot topic in condensed matter physics~\cite{Holzhey,Vidal,Calabrese,Plenio,Wolf,Barthel,Riera,Li}.

The violation occurs in cases of one-dimensional (1D) critical systems and models with Fermi surface. In these cases, the scaling contains logarithmic correction,
\begin{eqnarray}
S_{L}=\frac{1}{3}C L^{d-1}\log L, \label{CFT}
\end{eqnarray}
where $C$ is related to the number of excitation modes across the boundary between $A$ and $B$, and is equal to the central charge $c$ of conformal field theory (CFT) in $d=1$. In the CFT, the entropy is roughly given by a logarithm of a two-point correlation function for scaling operators, and thus Eq.~(\ref{CFT}) naturally appears. Away from a critical point, the entropy is deformed as
\begin{eqnarray}
S=\frac{1}{6}c{\cal A}\log\xi,\label{deformed}
\end{eqnarray}
with correlation length $\xi$ and the number of boundary points ${\cal A}$ of $A$.

On the other hand, there is another type of entropy scaling which does not contain the universality parameter $c$ and any length scale explicitely. When we take 1D quantum critical models by using variational optimization of matrix product state (MPS)~\cite{Ostlund}, it is conjectured that the half-chain entanglement entropy behaves as
\begin{eqnarray}
S_{\chi}=\frac{1}{6}\log\chi, \label{mps}
\end{eqnarray}
where $\chi$ is matrix dimension of MPS. This conjecture was recently found in two specific models with different central charges, respectively: one is transverse-field Ising chain ($c=1/2$), and the other is $S=1$ $XXZ$ chain with uniaxial anisotropy ($c=1$)~\cite{Tagliacozzo,Huang}. The $\log\chi$ dependence on this entropy can be interpleted as a result of quantum entanglement between $A$ and $B$. This is because $S_{\chi}=\log\chi$ for the maximally entangled-pair state, $\left|\psi\right>=(1/\sqrt{\chi})\sum_{n=1}^{\chi}\left| n\bar{n}\right>$, where the states in $A$ and $B$ are labeled by $n$ and $\bar{n}$, respectively, and one of the two degrees of freedom, $n$ or $\bar{n}$, is traced out. Here, $\left|\psi\right>$ is also a particular form of MPS. Furthermore, a prefactor $1/6$ is expected to be a character of the Virasoro algebra in CFT, although the microscopic understanding has not been obtained yet.

In general, MPS for $\chi=1$ represents local approximation (no entanglement, $S_{1}=0$), and is asymptotically exact if we could take a sufficiently large $\chi$ value. Then, approximately taking a finite $\chi$ value would limit spin correlation or quantum entanglement to finite-range one. In that sense, $\chi$ controls the range of the spin correlation. By combining Eq.~(\ref{mps}) with Eq.~(\ref{deformed}), we know that the effective correlation length of MPS is given by
\begin{eqnarray}
\xi_{\rm eff}=\chi^{1/c}, \label{issue}
\end{eqnarray}
where we take ${\cal A}=1$ because Eq.~(\ref{mps}) is obtained for the half of an infinitely-long chain. Therefore, this $\chi$ value is actually related to the length scale that represents how presicely the spin correlation is taken into account. However, this is somehow strange, since $\chi$ is just a parameter for how many singular values of the matrices in MPS are taken. The above consideration suggests that the length-scale control given by Eq.~(\ref{issue}) is a fundamental function of the singular value decomposition (SVD). The issue to be resolved here is why the SVD automatically produces the length scale.

Here, we address this issue in a viewpoint of quantum / classical correspondence. Usually, a 1D quantum spin model is transformed into a 2D classical spin model by the Suzuki-Trotter decomposition. Then, we can handle a Monte Carlo (MC) simulation, and obtain a snapshot of particular spin configuration. The correspondence may predict that a length scale characterized by $\chi$ in the quantum side is hidden in the snapshot. Therefore, we attempt to search the hidden length scale, and discuss about physical meaning of Eq.~(\ref{issue}). This is a purpose of this paper.

For this purpose, we introduce new entanglement entropy for a snapshot calculated by a MC simulation. This is the von Neumann entropy defined by the singular values of the reduced density matrix for the snapshot. Then, we find two scaling relations of the entropy in the Ising and the $3$-state Potts models that are analogous to Eq.~(\ref{deformed}) (or Eq.~(\ref{CFT})) and Eq.~(\ref{mps}). Furthermore, the scaling also appears on the positions of the multiple gaps in the entanglement spectrum. A key factor for the scaling is fractal spin configuration at $T_{c}$. The two scaling relations come from short- and long-range spin correlation in the fractal. A role of the SVD on the length-scale control is to decompose the original snapshot into a set of images with different length scales, respectively. Then, each scale is characterized by one of the multiple gaps in the entanglement spectrum. We discuss about similarity and difference between the new entropy scaling and standard one in 1D quantum systems, and also discuss about possible presence of a topological term hidden in our scaling relation.

The paper is organized as follows. In Sec.~II, we define the entanglement entropy for a snapshot, which is a key ingredient in this paper, and present outline of our method. Then, in Sec.~III, basic properties of the entanglment entropy and the entanglement spectrum for square-lattice Ising ferromagnet are presented. The main objective is to show temperature and system-size dependence of the entropy as well as the spectrum in order to extract scaling relations. We also examine entanglement support of our method by changing the number of the singular values which are taken into account. In Sec.~IV, coarse-grained snapshots are shown, and we discuss about the key mechanism of the length-scale control hidden in the SVD. In Sec.~V, we examine the $3$-state Potts model in order to confirm universality of our scaling relations obtained in the analysis of the Ising model. In Sec.~VI, we discuss about the topological entanglement entropy in a viewpoint of the entanglement gap. Finally, we summarize our results.

\section{Method}

We start with the Ising model on the square lattice:
\begin{eqnarray}
H=-J\sum_{\left<i,j\right>}\sigma_{i}\sigma_{j}.
\end{eqnarray}
where $\sigma_{i}=\pm 1$ and the sum runs over the nearest neighbor lattice sites $\left<i,j\right>$, and $J (>0)$ is exchange interaction. The system size is taken to be $L\times L$. According to dual transformation, the critical temperature is known to be $T_{c}/J=2/\log (1+\sqrt{2})=2.2692$. The central charge of the Ising model is $c=1/2$.

First, we are going to obtain a snapshot of a spin configuration $m(x,y)=\sigma_{i}$ with $i=(x,y)$. We can freely choose a method for obtaining the snapshot. Here, we will use MC simulation. We regard $m(x,y)$ as a matrix, and calculate the reduced density matrices defined by
\begin{eqnarray}
\rho_{X}(x,x^{\prime})&=&\sum_{y}m(x,y)m(x^{\prime},y), \label{densitymatrix1} \\
\rho_{Y}(y,y^{\prime})&=&\sum_{x}m(x,y)m(x,y^{\prime}), \label{densitymatrix2}
\end{eqnarray}
where we trace over $y$ ($x$)-component in $\rho_{X}$ ($\rho_{X}$). Let us decompose $m(x,y)$ into a set of the sigular values $\{\Lambda_{n}\}$ and the column unitary matrices $\{U_{n}(x)\}$ and $\{V_{n}(y)\}$:
\begin{eqnarray}
m(x,y)=\sum_{n=1}^{L}U_{n}(x)\sqrt{\Lambda_{n}}V_{n}(y). \label{SVD}
\end{eqnarray}
Mathematically, $\{\Lambda_{n}\}$ are uniquely determined, while $\{U_{n}\}$ and $\{V_{n}\}$ are not. Thus if some universal features could be extracted from a snapshot, those should be represented by a function of $\Lambda_{n}$. By substituting Eq.~(\ref{SVD}) into Eqs.~(\ref{densitymatrix1}) and (\ref{densitymatrix2}), we have
\begin{eqnarray}
\rho_{X}(x,x^{\prime})&=&\sum_{n=1}^{L}U_{n}(x)\Lambda_{n}U_{n}(x^{\prime}), \\
\rho_{Y}(y,y^{\prime})&=&\sum_{n=1}^{L}V_{n}(y)\Lambda_{n}V_{n}(y^{\prime}).
\end{eqnarray}
Thus, the set of $\{\Lambda_{n}\}$ is obtained by diagonalizing $\rho_{X}$ or $\rho_{Y}$. It is noted that the eigenvalue spectrum of $\rho_{X}$ is the same as that of $\rho_{Y}$. Even if we consider a rectangular lattice with $M\times N$ sites, the nonzero eigenvalues of $\rho_{X}$ and $\rho_{Y}$ are the same, and the number the eigenvalues is $L={\rm min}(M,N)$. We align the eigenvalues so that $\Lambda_{1}\ge\Lambda_{2}\ge\cdots\ge\Lambda_{L}$. Each eigenvalue $\Lambda_{n}$ is normalized to be $\lambda_{n}=\Lambda_{n}/C$ with a constant $C$ so as to satisfy \begin{eqnarray}
\sum_{n=1}^{L}\lambda_{n}=1.
\end{eqnarray}

Next, we define the von Neumann entropy of a snapshot analogous to Eq.~(\ref{EE}). That is the amount of entanglement between $x$- and $y$-components defined by
\begin{eqnarray}
S_{\chi}=-\sum_{n=1}^{\chi}\lambda_{n}\log\lambda_{n}, \label{entropy}
\end{eqnarray}
with $\chi\le L$. We abbreviate $S_{L}$ to $S$. This is a key quantity in this study. However, at the present stage, we do not know whether this is related to the standard entanglement entropy shown in Eqs.~(\ref{CFT}) and (\ref{mps}). Hereafter, we will present basic properties of $S$ in detail. Before going to the detail, it is theoretically clear that this entropy becomes maximum when we take $\lambda_{n}=1/\chi$ for any $n$. Then we have
\begin{eqnarray}
S_{\chi}\le -\sum_{n=1}^{\chi}\frac{1}{\chi}\log\frac{1}{\chi}=\log\chi,
\end{eqnarray}
and $S_{\chi}$ is bounded by $\log\chi$.

I have performed MC simulation by a standard Metropolis algorithm in order to obtain $m(x,y)$. Periodic boundary condition is taken into account for the square lattice. Starting with the temperature $T=3.02J$, $10^{6}$ MC steps are taken for thermal equillibrium. Here, one MC step counts $L\times L$ updates. After that, I gradually reduce temperature by $\Delta T=0.05J$ and take $10^{5}$ MC steps for convergence at each $T$. When calculating temperature dependence of the total entropy $S$ in detail, I take $\Delta T=0.01J$, and in this case the MC step in each $T$ is taken to be $10^{4}\sim 10^{5}$ depending on the system size. I have also confirmed numerical convergency by taking $10^{6}$ MC steps for some $T$ values. I have observed snapshots and their entropy across $T_{c}$. 

\section{Entanglement entropy and entanglement spectrum of the Ising model}

\subsection{Scaling relation for the entropy at $T\gtsim T_{c}$}

\begin{figure}[htbp]
\begin{center}
\includegraphics[width=9cm]{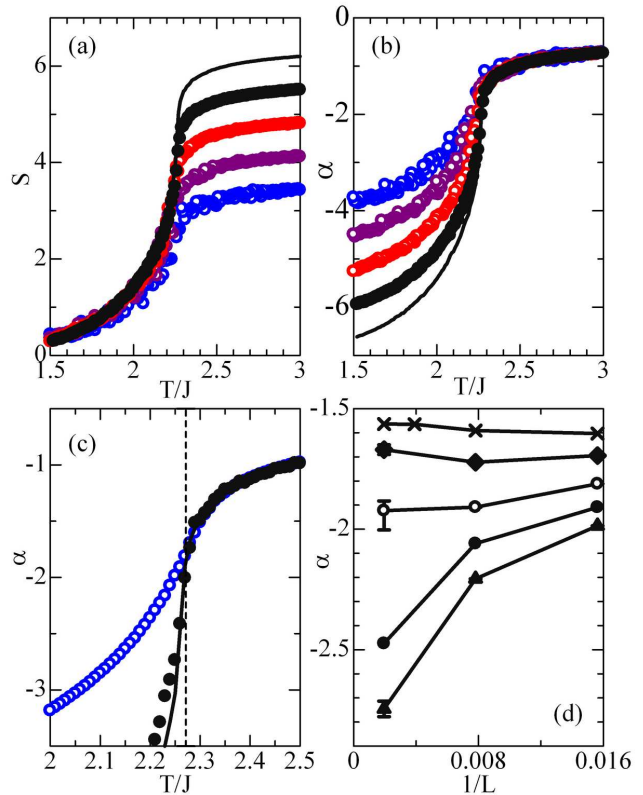}
\end{center}
\caption{(a-c) Temperature and system-size dependence on the entropy $S$ and $\alpha=S-\log L$: $L=64=2^{6}$ (blue circles), $L=128=2^{7}$ (purple circles), $L=256=2^{8}$ (red circles), $L=512=2^{9}$ (black circles), $L=1024=2^{10}$ (solid line). In pannel (c), we have avaraged $10^{6}$ samples for $L=64$. A dashed vertical line is a guide to $T_{c}$. (d) finite-size scaling for $\alpha$ near $T_{c}$: $T=2.25J$ (filled triangles), $T=2.26J$ (filled circles), $T=2.27J\sim T_{c}$ (open circles), $T=2.28J$ (filled diamonds), and $T=2.29J$ (crosses). We have avaraged $10^{6}$ and $10^{4}$ samples for $L\le 256$ and $L=512$, respectively. Note that a relative statistical error between $10^{4}$ and $10^{6}$ samples is $\Delta\alpha\sim 0.04$ for $L=256$ at $T=2.29J$.}
\label{fig1}
\end{figure}

Let us look at Fig.~\ref{fig1} where basic properties of the entropy $S$ are summarized. Since we would like to examine the amount of information in one snapshot, we do not take statistical average of $S$ except for cases that we need precise scaling. Fortunately, the statistical error of $S$ becomes smaller as $L$ increases as shown in Figs.~\ref{fig1}~(a) and (b), and the effect of self-avarage on $S$ seem to be much better than that of thermodynamic quantities. This small variance guarantees that our data do not suffer from the severe statistical error at least for large-$L$ region.

As for $T_{c}$, temperature dependence of $S$ is a good measure, since $S$ behaves quite differently below and above $T_{c}$ as shown in Fig.~\ref{fig1}~(a). Below $T_{c}$, the system is in the ferromagnetically ordered state ($\lambda_{1}=1$ and otherwise $0$), and then $S$ should go to zero. Above $T_{c}$, the spin configuration is paramagnetic, leading to high entropy. In that sense, the $T$-dependence is similar to that of the thermal entropy. We see that $S$ for $L=1024$ largely drops at $2.26J\le T\le 2.27J$ with decreasing temperature, suggesting phase transition. This position is very close to the exact $T_{c}$. Furthermore, at $T\ge T_{c}$, we find
\begin{eqnarray}
S = \log L + \alpha, \label{pseudoCFT}
\end{eqnarray}
with $\alpha<0$. Figure~\ref{fig1}~(b) plots $\alpha=S-\log L$ instead of $S$ in order to show this scaling clearly. In the next paragraph, we will obtain $\alpha=-\pi/4\sim -0.77$ in the large-$T$ and the large-$L$ limits. Actually, $\alpha$ at $T=3.02J$ is close to $-\pi/4$. At $T_{c}$, the value is estimated to be $\alpha\sim -2$ as shown in Fig.~\ref{fig1}~(c), where the data for $L=64$ are avaraged by $10^{6}$ samples. We pick up each sample per $1$ MC step after thermalization. The finite-size scaling for $\alpha$ near $T_{c}$ is also presented in Fig.~\ref{fig1}~(d), and the result supports $\alpha\sim -2$.

However, the $\log L$ dependence on $S$ at and above $T_{c}$ seems to come from different origins. In Fig.~\ref{fig1}~(d), the $\alpha$ value above $T_{c}$ ($T=2.29J$) increases slightly with $L$, and finally converges for $L\gtsim 512$. On the other hand, the $\alpha$ value gradually decreases with increasing $L$ at $T=2.27J$. Furthermore, the eigenvalue distribution has strong $T$-dependence particularly near $T_{c}$ as shown later. In the following subsections, we examine the physical origins of the scaling Eq.~(\ref{pseudoCFT}) at and above $T_{c}$ separately.

\subsection{Random matrix theory in the large-$T$ limit}

\begin{figure}[htbp]
\begin{center}
\includegraphics[width=9cm]{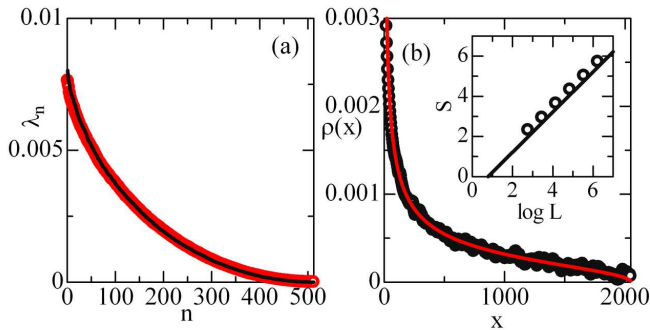}
\end{center}
\caption{(a) Normalized eigenvalues $\lambda_{n}$ ($L=512$): $T=10.0J$ (black) and $T=40J$ (red circles). (b) Eigenvalue distribution $\rho(x)$ ($L=512$): $T=10J$ (black circles). A red solid line represents the asymptotic distribution curve by the Mar$\check{\rm c}$enko-Pastur law. A broadening factor in Eq.~(\ref{dos2}) is taken to be $\gamma=10\ll\Lambda^{max}\sim 4L$. The inset shows the numerically obtained entropy $S$ as a function of $\log L$ at $T=10J$. The MC steps are $10^{4}$. A guide line in the inset represents Eq.~(\ref{logl}).}
\label{fig2}
\end{figure}

Let us first examine the eigenvalues above $T_{c}$. In Fig.~\ref{fig2}~(a), the eigenvalue $\lambda_{n}$ as a function of $n$ decays slowly, and then all of the eigenvalues play a role on the entropy. This behavior is unchanged for large-$T$ region, and we see that the data with $T=10J$ and $40J$ are almost the same. In the large-$T$ limit, the upper bound of $S$ is precisely determined by random matrix theory, since the spin configuration is paramagnetic (random). Here, we introduce the eigenvalue distribution
\begin{eqnarray}
\rho(x)&=&\frac{1}{L}\sum_{n=1}^{L}\delta(x-\Lambda_{n}) \label{dos} \\
&=&\lim_{\gamma\rightarrow 0+}\frac{1}{L}\sum_{n=1}^{L}\frac{1}{\pi}\frac{\gamma}{(x-\Lambda_{n})^{2}+\gamma^{2}}, \label{dos2}
\end{eqnarray}
and according to the random matrix theory $\rho(x)$ should asymptotically approach the Mar$\check{\rm c}$enko-Pastur law in the large-$L$ limit
\begin{eqnarray}
\rho(x)=\frac{1}{2\pi L x}\sqrt{x(4L-x)}, \label{MP}
\end{eqnarray}
for $0<x<4L$ with variance $L$. Actually, the numerically obtained distribution for $T=10J$ and $L=512$ well fit with this equation as shown in Fig.~\ref{fig2}~(b). For those parameters, the average of off-diagonal components of $\rho_{X}$ is $0.2662$ which is very small, and the variance of the off-diagonal components is $530.07\sim L$. The maximum eigenvalue is $\Lambda^{max}=2078.125\sim 4L$. These data also fit with the random matrix theory. Then, $S$ can be evaluated as follows: We transform Eq.~(\ref{entropy}) with $\chi=L$ into an integral form with use of Eq.~(\ref{dos})
\begin{eqnarray}
S&=&-\sum_{n=1}^{L}\frac{\Lambda_{n}}{C}\log\left(\frac{\Lambda_{n}}{C}\right) \nonumber \\
&=&-\int_{0}^{4L}dx\sum_{n=1}^{L}\delta(x-\Lambda_{n})\frac{x}{C}\log\left(\frac{x}{C}\right) \nonumber \\
&=&-\int_{0}^{4L}dx L\rho(x)\frac{x}{C}\log\left(\frac{x}{C}\right),
\end{eqnarray}
and substituting Eq.(\ref{MP}) to this equation. Then, we obtain the following scaling relation:
\begin{eqnarray}
S&=&-\int_{0}^{4L}dx\frac{\sqrt{x(4L-x)}}{2\pi L^{2}}\log\left(\frac{x}{L^{2}}\right) \nonumber \\
&=&\log L -\int_{-1}^{1}dt\sqrt{1-t^{2}}\log\left(2(1+t)\right) \nonumber \\
&=&\log L - \frac{\pi}{4}, \label{logl}
\end{eqnarray}
where $C$ is taken to satisfy
\begin{eqnarray}
C=\sum_{n=1}^{L}\Lambda_{n}=\int_{0}^{4L}dx\rho(x) x=L^{2}.
\end{eqnarray}
We have confirmed that this relation is strictly hold in our numerical simulation. Therefore, the $\log L$ dependence in the large-$T$ limit comes from the universal feature of the random matrix, and the diviation from the completely random state is characterized by $\alpha$. Numerical data in the inset of Fig.~\ref{fig2}~(b) support this analytical result.

\subsection{Eigenvalue distribution and Entanglement spectrum at $T_{c}$}

\begin{figure}[htbp]
\begin{center}
\includegraphics[width=9cm]{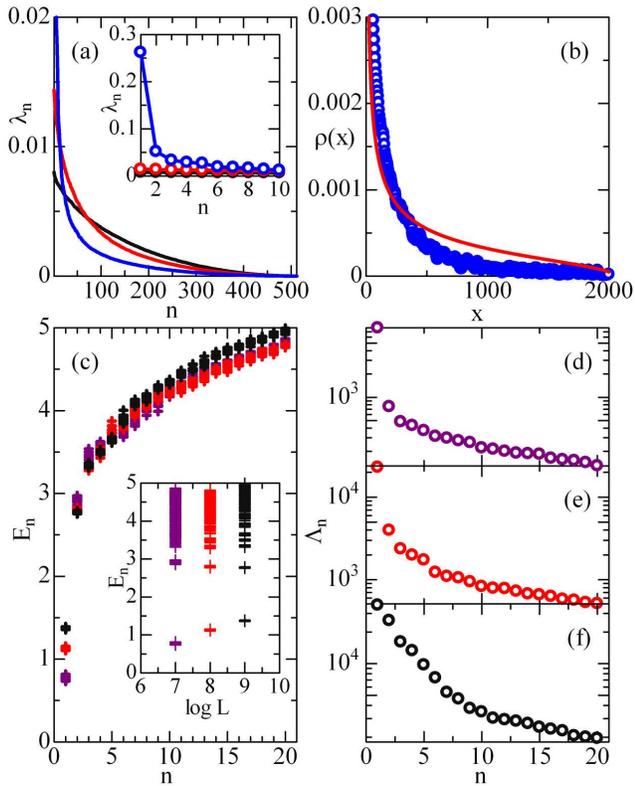}
\end{center}
\caption{(a) Normalized eigenvalues $\lambda_{n}$ ($L=512$): $T=2.27J$ (blue), $T=3.02J$ (red), and $T=10.0J$ (black). The inset is $\lambda_{n}$ for small $n$ region. (b) Eigenvalue distribution $\rho(x)$ ($L=512$): $T=2.27J$ (black open circles). A red solid line represents the asymptotic distribution curve by the Mar$\check{\rm c}$enko-Pastur law. We take $\gamma=10$. (c) Entanglement spectrum for $T=2.27J$: $L=128$ (red), $L=256$ (purple), and $L=512$ (black). We plot $10$ different-sample data for each parameter on the same pannel. (d-f) Logarithmic plot of eigenvalue $\Lambda_{n}$ for $L=128$, $256$, and $512$.}
\label{fig3}
\end{figure}

On the other hand, the nature of the eigenvalues at $T_{c}$ is quite different from that above $T_{c}$. We present the eigenvalues and the distribution function in Figs.~\ref{fig3} (a) and (b). In Fig.~\ref{fig3}~(a), we observe evolution of $\lambda_{n}$ in small $n$ region toward $T_{c}$, and finally $\lambda_{1}\rightarrow 1$ at zero temperature. Thus, the data at $T_{c}$ can be viewed as a mixture of various length scales, and small and large $n$ regions represent long-range (ferromagnetic) and short-range (paramagnetic) components of spin correlation, respectively. As shown in Fig.~\ref{fig3}~(b), the eigenvalue distribution deviates from that of the random matrix, and thus the $\log L$ dependence at $T_{c}$ is not due to the random matrix. Here, the average and the variance of off-diagonal components of $\rho_{X}$ are $92.022\gg 0$ and $5612.122 \gg L$, respectively, and $\Lambda^{max}=53350.01 \gg 4L$. These data also differ from those expected by the random matrix theory.

The separation of $\lambda_{n}$ into different length scales is more clearly seen in the entanglement spectrum~\cite{Calabrese2} defined by
\begin{eqnarray}
E_{n}=-\log\lambda_{n}.
\end{eqnarray}
In Fig.~\ref{fig3}~(c) and the inset, we plot $10$ different-sample data for each parameter in order to reduce finite-size effects. We find multiple entanglement gaps near $T_{c}$~\cite{Thomale}. In addition to large gaps at around $E_{n}=2$ and $3$, gap-like anomalies also appear at $E_{n}\sim 3.75$ and $4.25$ for $L=512$. Because of the presence of the multiple gaps, the states are separated into a set of different entanglement levels. The state below the largest gap represents a precursor of ferromagnetic long-range order, while nearly continuum states far above the large gaps represent paramagnetic short-range spin configuration. In between, each sector separated by the gaps would have intermediate length scales of spin correlation. Then, we expect that the multiple gaps characterize various length scales, and their mixture leads to critical behavior.

Let us look at more about the gap-like anomalies above $E_{n}=3$ in Figs.~\ref{fig3}~(d-f), where $\log_{10}\Lambda_{n}\propto E_{n}$ is plotted. We find a kink structure, and the position of the kink, $n_{kink}$, exactly traces one of the multiple entanglement gaps. The kink position increases as $L$. We observe that $n_{kink}=3$, $6$, and $9$ for $L=2^{7}$, $2^{8}$, and $2^{9}$, respectively. Thus, this observation tells us
\begin{eqnarray}
\frac{1}{3}n_{kink}=\log_{2}L-6, \label{kink}
\end{eqnarray}
for $L> 2^{6}$. The kink position gives us a criterion of how many eigenvalues for the given $L$ play an important role on the spin correlation associated with the critical behavior. As we increase $L$, the discreteness of the lattice tends to disappear, and eventually various sizes of ferromagnetic clusters appear at $T_{c}$. Therefore, $n_{kink}$ increases as $L$. The kink structure does not appear in Fig.~\ref{fig2}~(a) for large $T$, and thus Eq.~(\ref{kink}) is a character which only appears near $T_{c}$. We consider that this $\log L$ dependence on $n_{kink}$ is closely related to the $\log L$ dependence on $S$ in Eq.~(\ref{pseudoCFT}). It should be noted that only few states are concerned with the long-range spin correlation of the critical behavior in our expression. This is quite contrast to the MPS scaling, and this point will be important in the later discussion.

\subsection{Compasiron with the thermal entropy}

\begin{figure}[htbp]
\begin{center}
\includegraphics[width=5cm]{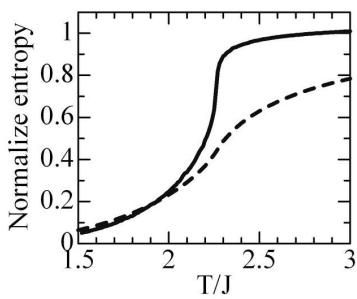}
\end{center}
\caption{Comparison between $S/(\log L -\pi/4)$ (solid line, $L=1024$) with $s^{thermal}/\log 2$ (dashed line, exact).}
\label{fig4}
\end{figure}

In order to see the pecuriality of $S$ near $T_{c}$, it is meaningful to compare $S$ with the thermal entropy $S^{thermal}$. The thermal entropy also goes to zero below $T_{c}$ due to ferromagnetic order, while in the large-$T$ limit we have a high constant value $L^{2}\log 2$ coming from the number of all possible spin configurations $2^{L^{2}}$. Thus, the physical meaning of $S$ and $S^{thermal}$ is essentially the same in the both limits. Figure~\ref{fig4} shows comparison between $S$ and $S^{thermal}$. The exact form of $S^{thermal}$ per site in the thermodynamic limit, $s^{thermal}=\lim_{L\rightarrow\infty}S^{thermal}/L^{2}$, is given by the Onsager's solution. First, we transform the thermodynamic first law into the following form
\begin{eqnarray}
s^{thermal}=-\beta\frac{\partial}{\partial\beta}(-\beta f)-\beta f,
\end{eqnarray}
with free energy per site $f$ and inverse temperature $\beta=1/T$. Then we substitute the Onsager's free energy into the above equation to obtain $s^{thermal}$:
\begin{eqnarray}
-\beta f=\frac{1}{2}\log(2\sinh 2K)+\frac{1}{2\pi}\int_{0}^{\pi}dq\epsilon(q,K),
\end{eqnarray}
where $K=\beta J$, and $\epsilon(q,K)$ is a solution of the following equation
\begin{eqnarray}\cosh\epsilon(q,K)=\cosh 2K\coth 2K-\cos q.
\end{eqnarray}

In Fig.~\ref{fig4}, we show the normalized data so that both of them approach unity in the large-$T$ limit. We find that the change in $S$ near $T_{c}$ is more cusp-like in comparison with $S^{thermal}$. This cusp-like feature is an evidence that separates the $\log L$ dependence near $T_{c}$ from that above $T_{c}$. The cusp reminds us a fact that the entanglement entropy in transverse-field Ising model increases toward the quantum critical point.

\subsection{Finite-$\chi$ scaling}

\begin{figure}[htbp]
\begin{center}
\includegraphics[width=8.5cm]{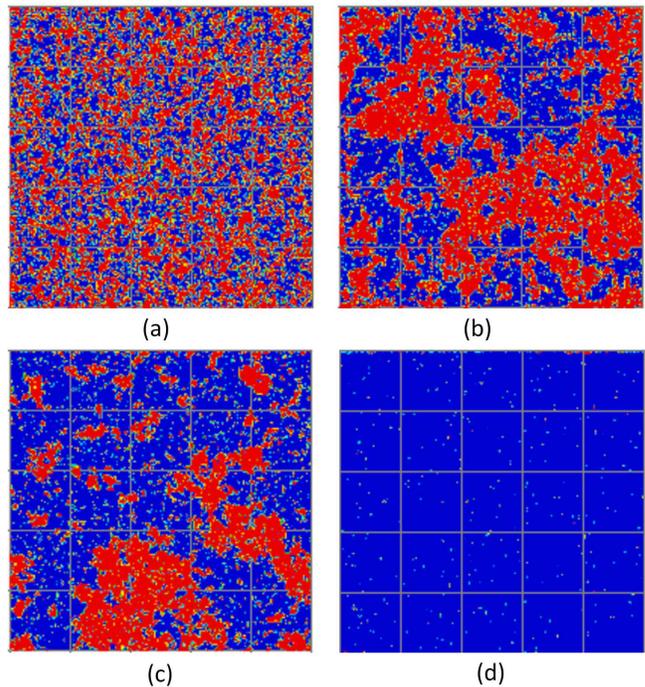}
\end{center}
\caption{Snapshots of spin configurations ($L=256$): (a) $T/J=3.02$, (b) $T/J=2.32$, (c) $T/J=2.27$, (d) $T/J=1.52$.}
\label{fig5}
\end{figure}

\begin{figure}[htbp]
\begin{center}
\includegraphics[width=8.5cm]{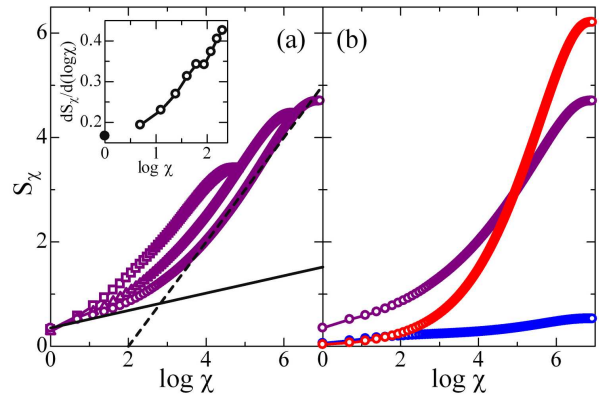}
\end{center}
\caption{(a) $S_{\chi}$ for $L=1024$ at $T=2.27J\sim T_{c}$ (purple circles). For compasiron, we also plot the data for $L=128$ (purple box) and $L=512$ (purple triangles) at $T=2.27J$. A solid line with slope $1/6$ is a guide to the eye. A dashed line represents $S_{\chi}=\log\chi-2$. The inset shows $\partial S_{\chi}/\partial\log\chi$ for $L=1024$. The black filled circle represents $1/6$. (b) Temperature dependence of $S_{\chi}$ ($L=1024$): $T=3.02J$ (red) and $T=1.52J$ (blue).}
\label{fig6}
\end{figure}

The scaling Eq.~(\ref{pseudoCFT}) at $T_{c}$ can be understood by calculating $\chi$ dependence on $S_{\chi}$. Figures~\ref{fig5} and \ref{fig6} show snapshots of spin configurations and their entropy $S_{\chi}$ at various $T$. We take $L=1024$, since we know that $\alpha$ is almost converged for this size as discussed in Fig.~\ref{fig1}. We find that the entropy asymptotically approaches
\begin{eqnarray}
S_{\chi}=\frac{1}{6}\log\chi + \gamma^{\prime}, \label{scaling2}
\end{eqnarray}
for $\chi\ltsim n_{kink}$ as we increase $L$. The residual entropy $\gamma^{\prime}$ does not depend on $L$, and $\gamma^{\prime}=S_{1}\sim 0.35$. Now, we have $n_{kink}=12$ and $\log n_{kink}=2.485$ for $L=1024$. In the inset of Fig.~\ref{fig6}~(a), we plot the first derivative of $S_{\chi}$ with respect to $\log\chi$ in order to see the slope more precisely. Furthermore, for $\chi>n_{kink}$, we find
\begin{eqnarray}
S_{\chi}=\log\chi + \gamma \label{scaling},
\end{eqnarray}
with $\gamma=-2$ for $L=1024$. We rewrite Eq.~(\ref{scaling}) as
\begin{eqnarray}
S_{\chi} = \frac{1}{6}\log n_{kink}+\gamma^{\prime}+\Delta S_{\chi}.
\end{eqnarray}
As already mentioned, the additional term $\Delta S_{\chi}$ comes from short-range spin correlation. Since this component originates from paramagnetic spin configuration above $T_{c}$, the absolute value is larger than the $(1/6)\log\chi$ term.

Let us first look at Eq.~(\ref{scaling2}) in Fig.~\ref{fig6}~(b). In the ferromagnetically-ordered phase, the entropy should disappear, and actually the slope of $S_{\chi}$ at $T=1.52J<T_{c}$ is almost zero. As we approach $T=2.27J\sim T_{c}$, the slope of $S_{\chi}$ for $\chi\ltsim n_{kink}$ gradually increases toward $1/6$. However, the slope starts to decrease beyond $T_{c}$. Actually, the slope becomes very small for $T=3.02J>T_{c}$. According to the previous subsection, the slope near $T_{c}$ would be due to the long-range spin correlation.

Next we consider Eq.~(\ref{scaling}). We have already observed
\begin{eqnarray}
\gamma=-2=\alpha, \label{constant}
\end{eqnarray}
and then the scaling~(\ref{scaling}) after taking $\chi=L$ is considered to be the origin of Eq.~(\ref{pseudoCFT}) at $T_{c}$. In Fig~\ref{fig6}, $S_{\chi}$ somehow saturates at around $\chi=L$, but this is finite-size effect. We have numerically confirmed that the saturation reduces as we increase $L$. Thus, for $\chi\sim L$, Eq.~(\ref{scaling}) is strictly satisfied in the large-$L$ limit. This feature is also consistent with the previous result that $S$ asymptotically approaches Eq.~(\ref{pseudoCFT}) in the large-$L$ limit at $T_{c}$. With increasing $T$, $S_{\chi}$ is still proportional to $\log\chi$ for $\chi\gtsim n_{kink}$, while the slope gradually increases. At $T=3.02J>T_{c}$, the slope is about two times larger. Since the residual term $\gamma$ is a large negative value, $S_{\chi}$ finally approaches $\log L+\alpha$ at $\chi=L$.

All of the results presented in this section suggest that the origins of the $\log L$ dependence in Eq.~(\ref{pseudoCFT}) are clearly different at and above $T_{c}$.

\subsection{Block-spin transformation and scaling}

\begin{figure}[htbp]
\begin{center}
\includegraphics[width=5cm]{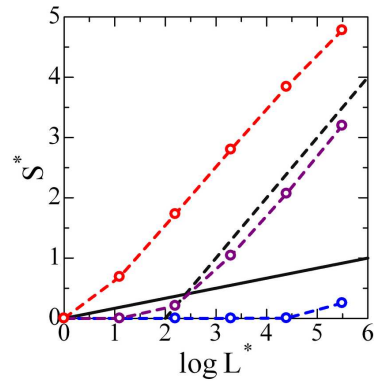}
\end{center}
\caption{Entanglement entropy $S^{\ast}$ for snapshots generated by the block-spin transformation at $T=3.02J$ (red), $T=2.27J$ (purple), and $T=1.52J$ (blue). The original snapshot is taken for $L=243=3^{5}$. We avarage $10^{4}$ samples for the calculation of $S^{\ast}$. Solid and dashed guide lines are $S^{\ast} = \log L^{\ast} - 2$ and $S^{\ast}=(1/6)\log L^{\ast}$, respectively.}
\label{fig7}
\end{figure}

In order to examine the physical meaning of Eqs.~(\ref{scaling2}-\ref{constant}), it is efficient to introduce the block-spin transformation and calculate the entanglement entropy $S^{\ast}$ associated with the transformation. The block-spin transformation merges $3\times 3$ lattice sites together, and the effective Ising spin on the new site (the number of the new sites is represented by $L^{\ast}$) also takes $1$ or $-1$ depending on spin configuration that more than half are up or down spins, respectively. We continue the transformation until the system becomes one effective site. Then, $S^{\ast}$ for $L^{\ast}\rightarrow 1$ represents the entropy of the fixed point of this renormalization group. The entropy $S^{\ast}$ is calculated by
\begin{eqnarray}
S^{\ast}=-\sum_{n=1}^{L^{\ast}}\lambda_{n}^{\ast}\log\lambda_{n}^{\ast}.
\end{eqnarray}
where $\lambda_{n}^{\ast}$ is the eigenvalue of the reduced density matrix in the coarse-grained system with the system size $L^{\ast}$.

Figure~\ref{fig7} shows $L^{\ast}$-dependence on $S^{\ast}$ at various $T$. At $T_{c}$, we again find the scaling
\begin{eqnarray}
S^{\ast}\sim\frac{1}{6}\log L^{\ast},
\end{eqnarray}
for small $L^{\ast}$ region, and
\begin{eqnarray}
S^{\ast}\sim\log L^{\ast} - 2,
\end{eqnarray}
for large-$L^{\ast}$ region. A crossing point of these lines is not far from $n_{kink}$. Thus, we see that $L^{\ast}$ plays a similar role on $\chi$ in Eqs.~(\ref{scaling2}) and (\ref{scaling}). Therefore, it is reasonable to consider that $\chi$ represents a length scale associated with the couarse graining. We also understand that the spin fluctuation in short-range scale has been already renormalized by the block-spin transformation in small $L^{\ast}$ region, and only large-scale phenomena survive. Then, Eq.~(\ref{scaling2}) characterizes this large-scale phenomena.

\section{Coarse-grained snapshot and length scale}

\subsection{Coarse-grained snapshot}

In order to understand the nature of the scaling relations in detail, it is important to remenber that ferromagnetic islands in the snapshot are fractal-like at $T_{c}$ (see Fig.~\ref{fig5}~(c)). Actually, the system is self-similar, and we can always observe various sizes of the islands even when we continue to zoom in the system. The variety of the island sizes plays a crucial role on the presence of the scaling relations. This should be also related to the presence of the multiple gaps in the entanglement spectrum. However, let us also remember the original definition of the entropy $S_{\chi}$, where $\chi$ is just the truncation number of the SVD and does not seem to connect directly to any length scale. Thus, we should examine how such a length scale is automatically generated by the SVD.

We again call SVD of the snapshot $m(x,y)=\sum_{n=1}^{L}U_{n}(x)\sqrt{\lambda_{n}}V_{n}(y)$. Our target is to look at a 'deformed' snapshot which is defined by restricting the sum upto $\chi$ in the SVD:
\begin{eqnarray}
m_{\chi}(x,y)=\sum_{n=1}^{\chi}U_{n}(x)\sqrt{\lambda_{n}}V_{n}(y),
\end{eqnarray}
where $m(x,y)=m_{L}(x,y)$. We expect that this should be a 'coarse-grained' image of the original snapshot, if $\chi$ characterizes a length scale.

\begin{figure}[htbp]
\begin{center}
\includegraphics[width=8.5cm]{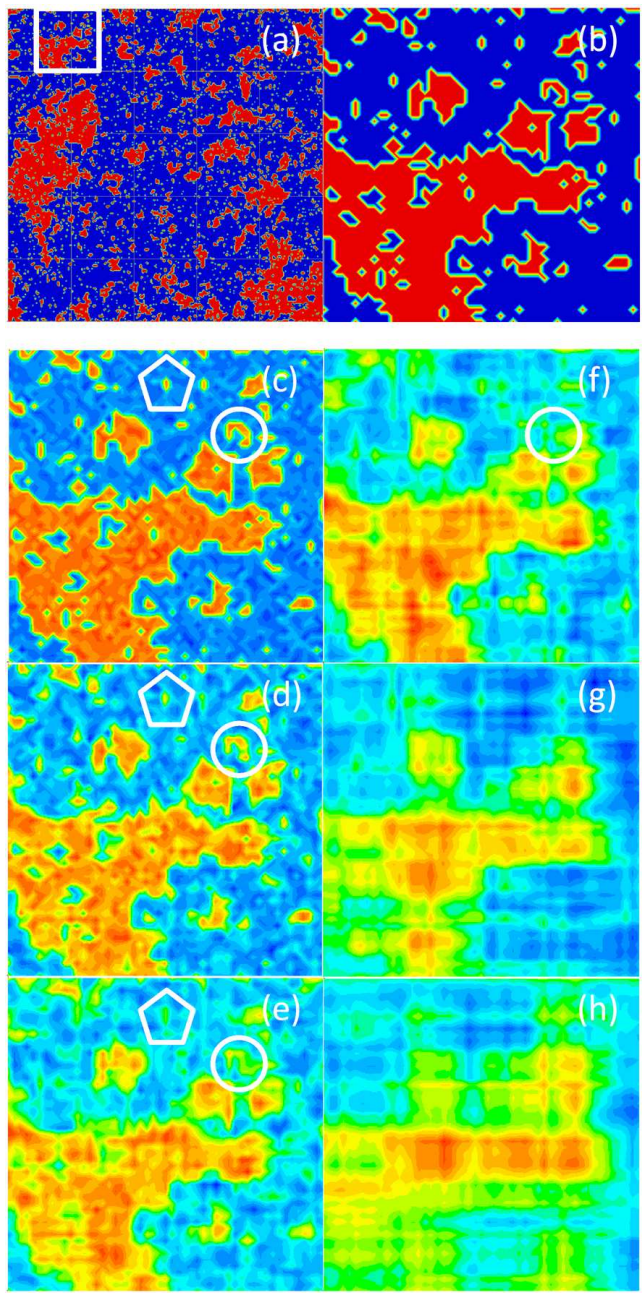}
\end{center}
\caption{Zooming in a region $25\le x\le 75$ and $206\le y\le 256$ of $m_{\chi}(x,y)$ for $T=2.27J$ and $L=256$: (a) original snapshot ($L=256$), (b) $\chi=L$, This image corresponds to white-square region of (a), (c) $\chi=L/2$, (d) $\chi=L/4$, (e) $\chi=L/8$, (f) $\chi=L/16$, (g) $\chi=L/32$, and (h) $\chi=L/64$.}
\label{coarsegraining}
\end{figure}

Figure~\ref{coarsegraining}~(a) is a target snapshot at $T=2.27J\sim T_{c}$, and Figs.~\ref{coarsegraining}~(b)-(h) zoom in a region $25\le x\le 75$ and $206\le y\le 256$ of $m_{\chi}(x,y)$ for various $\chi$. First, comparing Fig.~\ref{coarsegraining}~(a) with (b), we see that the original snapshot is fractal-like: even if we zoom in the snapshot, various sizes of ferromagnetic islands appear sequentially. Decreasing the value of $\chi$, we find that global structures do not change so much, but $m_{\chi}(x,y)$ gradually loses fine structures. The snapshot contains various sizes of the islands near $T_{c}$, and thus the lost of the fine structures means that larger and larger islands are damaged by the reduction of $\chi$. Let us look at one of the smallest islands (surrounded by a pentagon) that disappears in pannel (e). We also concentrate on the island surrounded by a circle. This island is a little bit larger than the smallest one. The island still remains in pannel (e), although the shape is deformed. Therefore, $\chi$ is actually controlling the accesible length scale of our model.

\subsection{Layered structure with different length scales}

\begin{figure}[htbp]
\begin{center}
\includegraphics[width=8cm]{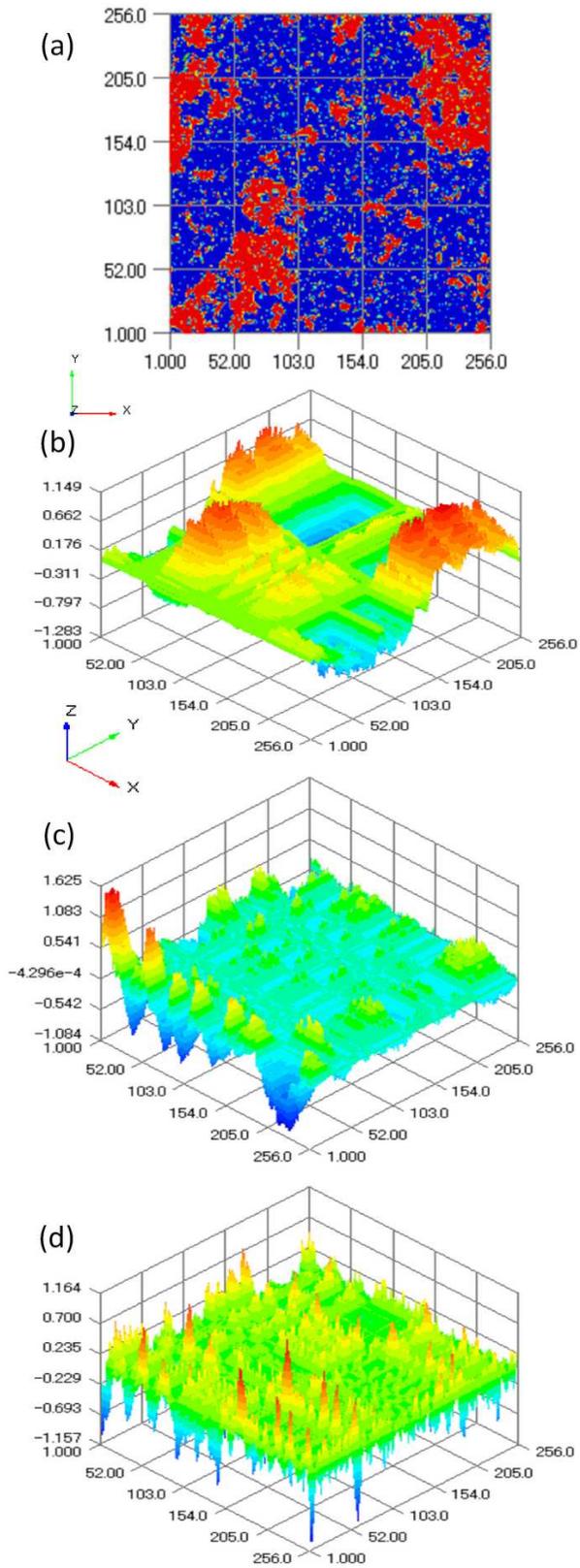}
\end{center}
\caption{Contour map of $m^{(n)}(x,y)$: $L=256$, $n_{kink}=6$, and $T=2.27J$, (a) original snapshot, (b) $n=2<n_{kink}$, (c) $n=4<n_{kink}$, and (d) $n=8>n_{kink}$.}
\label{fig9}
\end{figure}

\begin{figure}[htbp]
\begin{center}
\includegraphics[width=7.5cm]{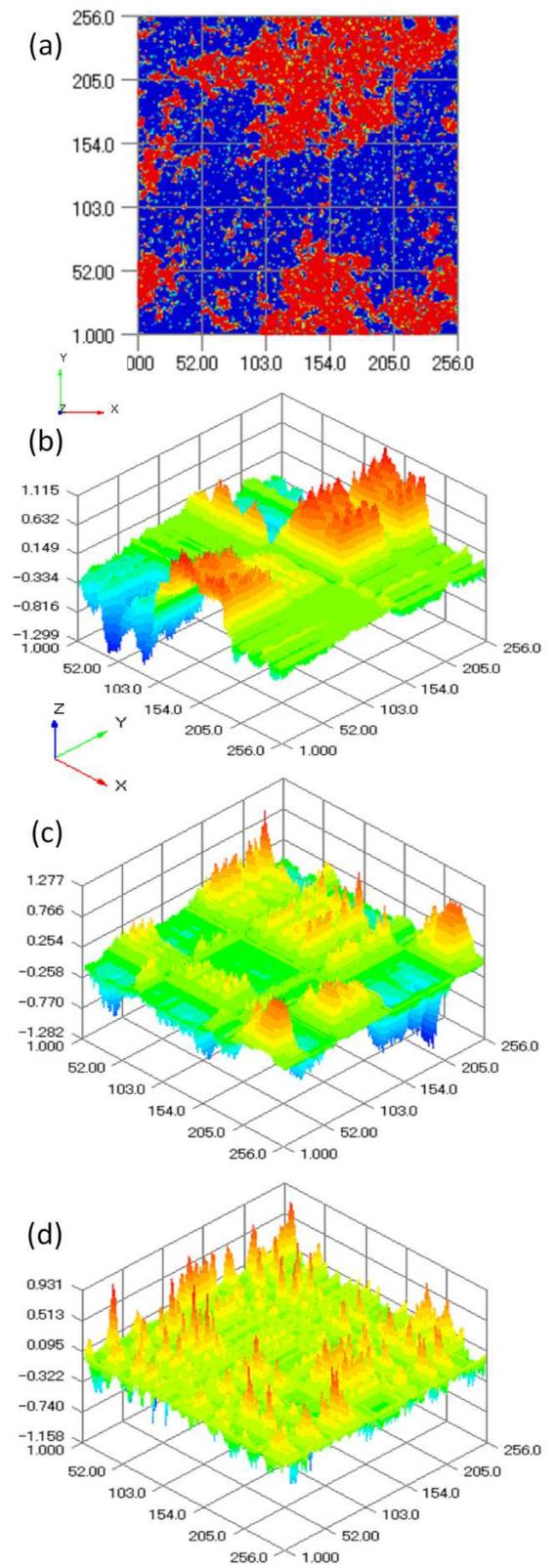}
\end{center}
\caption{Contour map of $m^{(n)}(x,y)$: $L=256$, $n_{kink}=6$, and $T=2.27J$, (a) original snapshot which differs from Fig.~\ref{fig9} (a), (b) $n=2<n_{kink}$, (c) $n=4<n_{kink}$, and (d) $n=8>n_{kink}$.}
\label{fig10}
\end{figure}

As already discussed, the length scale is characterized by the kink structure (multiple gap structure) in the entanglement spectrum. In order to see the effect of the structure on the snapshot, we present contour map for each layer of $m(x,y)$ defined by
\begin{eqnarray}
m^{(n)}(x,y)=U_{n}(x)\sqrt{\lambda_{n}}V_{n}(y),
\end{eqnarray}
where we call the label $n$ as 'layer'. The parameters are taken to be $L=256$ and $T=2.27J$. In Fig.~\ref{fig9}~(b) with $n=2$, we find that $m^{(2)}(x,y)\sim 1$ in wide $(x,y)$ region (see the vertical axis). When we compare Fig.~\ref{fig9}~(b) with (a), it is clear that the region represents the largest ferromagnetic islands. In Fig.~\ref{fig9}~(c) with $n=4$, we observe many broad peaks, and the positions of some of the representative peaks are assigned to be those of relatively large ferromagnetic islands. In Fig.~\ref{fig9}~(d) with $n=8$, the peaks become very sharp, and they represents small ferromagnetic islands. Further increasing the number of $n$, we can observe much shaper structures representing smaller and smaller islands. Since the spatial distribution of a single peak corresponds to the size of a ferromagnetic island, we understand that each layer has its own length scale. This is the reason why $\chi$ is a scaling parameter.

Let us remember that the kink structure in the entanglement spectrum appears at $n_{kink}=6$ for $L=256=2^{8}$. Then, we actually see that the data with $n=8>n_{kink}$ only represent local components. In the present case, a small $n$ represents large ferromagnetic islands, while in Eq.~(\ref{issue}) a large $\chi$ value represents successive treatment of long-range correlation. Thus, the role of $\chi$ on $\xi$ is reversed between quantum and classical sides. Figure~\ref{fig10} shows contour map for a snapshot with rather different island structures. The tendency of length-scale control also appear in this case. When we compare Fig.~\ref{fig9} with Fig.~\ref{fig10}, the average size of each island for a fixed $n$ value is similar in both Figures. This clearly shows the presence of the multiple entanglement gaps.

\subsection{Intuitive understanding of the length-scale control by SVD}

\begin{figure}[htbp]
\begin{center}
\includegraphics[width=6.5cm]{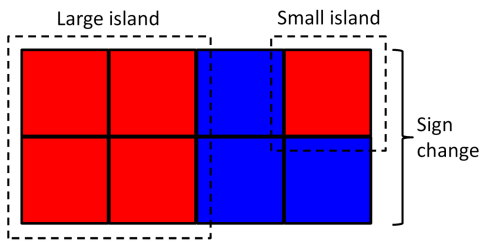}
\end{center}
\caption{$2\times 4$ lattice model for coexistence of small and large islands: $\sigma(x,y)=+1$ (red pixel) and $\sigma(x,y)=-1$ (blue pixel). The left half of the system represents a large ferromagnetic island, while $m(1,4)$ represents a small island. They are separated by background spins with $\sigma(x,y)=-1$. In this case, the area of the background spins are larger than that of the small island.}
\label{example}
\end{figure}

Let us explain the key mechanism of the length scale control hidden in SVD. Here, we introduce a snapshot with $2\times 4$ pixels shown in Fig.~\ref{example}. This is a minimal model of the snapshot where small and large ferromagnetic islands coexist. The magnetic moment $m$ of this snapshot is represented by the matrix
\begin{eqnarray}
m=\left(
\begin{array}{cccc}
1&1&-1&1 \\
1&1&-1&-1
\end{array}
\right).
\end{eqnarray}
Here, we call $m(1,4)$ as small ferromagnetic island, and call the left half as large ferromagnetic island. Then, the reduced density matrices $\rho_{X}$ and $\rho_{Y}$ are given by
\begin{eqnarray}
\rho_{X}&=&\left(
\begin{array}{cc}
4&2 \\
2&4
\end{array}
\right), \\
\rho_{Y}&=&\left(
\begin{array}{cccc}
2&2&-2&0 \\
2&2&-2&0 \\
-2&-2&2&0 \\
0&0&0&2
\end{array}
\right).
\end{eqnarray}
Their nonzero eigenvalues are the same, and they are $\lambda_{1}=6$ and $\lambda_{2}=2$. In the present ideal case, the density matrix $\rho_{Y}$ can be completely decoupled into two subspaces: $\rho_{Y}(4,4)$ represents the small island and $m(2,4)$, while the remaining $3\times 3$ submatrix represents the large island and two background spins located at $(x,y)=(1,3)$ and $(2,3)$.

The decomposition of $\rho_{Y}$ into two subspaces occurs due to rapid sign change of the magnetic moment at around the small island. The density matrix $\rho_{Y}$ is constructed by inner product between two column vectors of $m$, and then the inner product between the vectors for small and large islands vanishes due the sign change. Thus, the inner product means spatial correlation between the islands. As for the $3\times 3$ submatrix, the absolute values of the off-diagonal components are as large as the diagonal components. This is also clear, since the off-diagonal component comes from inner product between two column vectors inside of the large island. Then, the eigenvalues split into large one and zero, leading to $\lambda_{1}=6$. This situation is similar to that of the energy-level splitting into bonding and anti-bonding states after mixing of two orbitals.

By solving the eigenvalue equations, we obtain
\begin{eqnarray}
\vec{U}_{1}=\frac{1}{\sqrt{2}}\left(
\begin{array}{c}
1 \\
1
\end{array}
\right),
\vec{U}_{2}=\frac{1}{\sqrt{2}}\left(
\begin{array}{c}
1 \\
-1
\end{array}
\right),
\end{eqnarray}
and
\begin{eqnarray}
\vec{V}_{1}=\frac{1}{\sqrt{3}}\left(
\begin{array}{c}
1 \\
1 \\
-1 \\
0
\end{array}
\right),
\vec{V}_{2}=\left(
\begin{array}{c}
0 \\
0 \\
0 \\
1
\end{array}
\right).
\end{eqnarray}
Finally, each layer of $m=\sum_{n}m^{(n)}$ is reconstructed by
\begin{eqnarray}
m^{(1)}&=&\sqrt{6}\vec{U}_{1}\otimes{}^{t}\vec{V}_{1}=\left(
\begin{array}{cccc}
1&1&-1&0 \\
1&1&-1&0
\end{array}
\right), \\
m^{(2)}&=&\sqrt{2}\vec{U}_{2}\otimes{}^{t}\vec{V}_{2}=\left(
\begin{array}{cccc}
0&0&0&1 \\
0&0&0&-1
\end{array}
\right).
\end{eqnarray}
Therefore, we find that the small and large islands are automatically decoupled into different layers. Realistic cases are more complicated, but the feature of the rapid sign change still remains even in those cases.

\section{Three-state Potts model}

We would like to know whether our results are generalized for much broader universality classes. Here, we examine the $3$-state ($q=3$) Potts model:
\begin{eqnarray}
H=-J\sum_{\left<i,j\right>}\delta(\sigma_{i},\sigma_{j}),
\end{eqnarray}
with $\sigma_{j}=-1,0,1$, $\delta(\sigma,\sigma^{\prime})=1 (\sigma=\sigma^{\prime}), -1 (\sigma\ne\sigma^{\prime})$. The central charge for $Z_{q}$ symmetric CFT and the critical temperature are $c=2(q-1)/(q+2)=4/5$ and $T_{c}/J=2/\log\left(1+\sqrt{q}\right)=1.98994$. We perform MC simulation to obtain snapshots, where we take maximally $10^{7}$ MC steps. We remember that we need to take at least $L>64$ to catch the critical behavior in the Ising model.

\begin{figure}[htbp]
\begin{center}
\includegraphics[width=9cm]{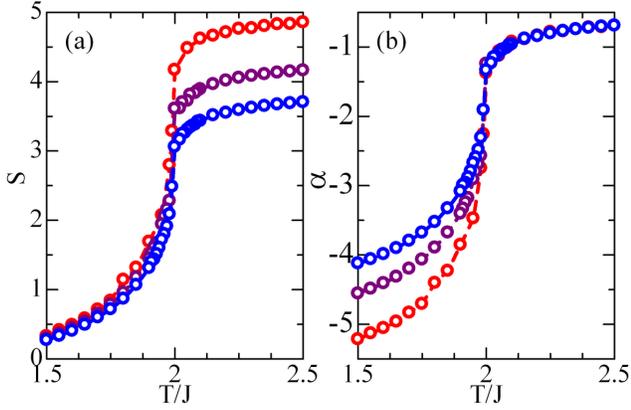}
\end{center}
\caption{(a)(b) $T$ and $L$-dependence of $S$ and $\alpha=S-\log L$ for $3$-state Potts model: $L=81=3^{4}$ (blue), $L=128=2^{7}$ (purple), and $L=256=2^{8}$ (red). The data for $L=81$ and $L=128$ are avaraged by $10^{4}$ samples. We do not take the avarage for $L=256$.}
\label{fig12}
\end{figure}

Figures~\ref{fig12}~(a) and (b) show $T$ and $L$-dependence of $S$ and $\alpha=S-\log L$. The abrupt decrease in $S$ occurs at around $T=1.99J$, suggesting that $T_{c}$ is very close to the exact value. In Fig.~\ref{fig12}~(b), we again find the scaling Eq.~(\ref{pseudoCFT}), $S=\log L+\alpha$, for $T\ge T_{c}$, and $\alpha=-\pi/4$ in the large-$T$ limit and $\alpha\sim -2$ at $T_{c}$. Therefore, Eq.~(\ref{pseudoCFT}) is hold in the $3$-state Potts model. The slope of the finite-$L$ scaling for the $\alpha$ value at $T_{c}$ is also different from that above $T_{c}$. The slope at $T_{c}$ decreases with $L$, while the slope above $T_{c}$ increase as $L$. The scaling at $T_{c}$, $S=\log L-2$, would come from critical behavior as we have already discussed in previous sections.

\begin{figure}[htbp]
\begin{center}
\includegraphics[width=9cm]{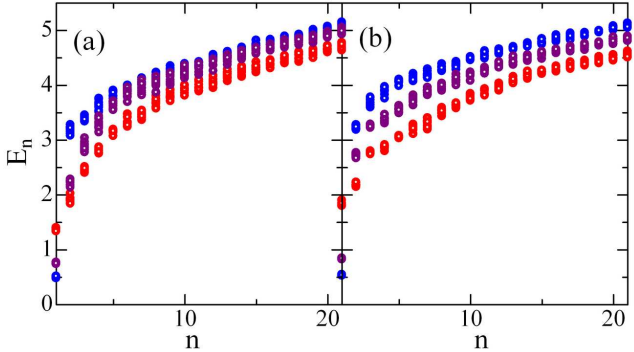}
\end{center}
\caption{Temperature dependence of entanglement spectrum $E_{n}$: $T=2.0J$ (red), $T=1.99J\sim T_{c}$ (purple), and $T=1.98J$ (blue). We plot the data of $10$ samples on the samle pannel. (a) $L=81=3^{4}$. (b) $L=128=2^{7}$.}
\label{fig13}
\end{figure}

Figure~\ref{fig13} shows $T$-dependence of $E_{n}$ for $L=81=3^{4}$ and $L=128=2^{7}$ near $T_{c}$. We find that the kink starts to appear with decreasing $T$, and finally the gap opens below $T_{c}$. Near the kink positions, ferromagnetic islands with two different length scales compete with each other, and then the data are somehow scattered. Thus, we use this scattering as a sign of the kink. The kink positions are located at $n_{kink}=3$ and $n_{kink}=8\sim 9$ for $L=81$ and $L=128$, respectively. The result suggests that $n_{kink}\propto\log_{3}L$, and actually the $\log L$ dependence on $\alpha$ at $T_{c}$ is related to critical behavior.

The data at $T=2.0J$ slightly above $T_{c}$ also bend at $n$ larger than $n_{kink}$. This is because $T$ is one of energy scales, and thus is related to the length scale. Higher temperature represents shorter wave length, and thus the bending position shifts to larger $n$ region as we increase $T$. Actually, in Fig.~\ref{fig13}~(b), a weak kink is observed at $n_{kink}=14\sim 15$. We expect that this kink becomes clear as we increase $L$.

\begin{figure}[htbp]
\begin{center}
\includegraphics[width=5cm]{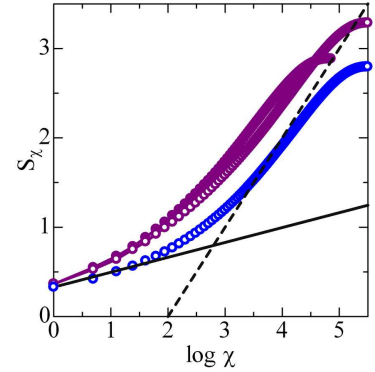}
\end{center}
\caption{$S_{\chi}$ at $T=1.99J\sim T_{c}$ for $L=256$ (open purple circles) and $L=128$ (filled purple circles). For comparison, $S_{\chi}$ at $T=1.98J$ is also plotted (blue circles). The slopes of the guide lines are $1/6$ and $1$.}
\label{fig14}
\end{figure}

We finally show finite-$\chi$ scaling for $S_{\chi}$ in Fig.~\ref{fig14}. We could not conclude definitely the presence of the scaling relations~(\ref{scaling2}) and (\ref{scaling}), but this size dependence looks similar to that in the Ising model. Therefore, we believe that the data asymptotically approach the scaling relations previously obtained, and the scaling, $S=\log L-2$, comes from $S_{\chi}=\log\chi-2$ for $\chi\rightarrow L$ in the large-$L$ limit.

\section{Discussion and Summary}

Up to now, we have observed two scaling relations concerned with length scales of ferromagnetic islands in the snapshot at $T_{c}$. The first one is given by
\begin{eqnarray}
S_{\chi}=\frac{1}{6}\log\chi+\gamma^{\prime}, \label{qc}
\end{eqnarray}
for $\chi\ltsim n_{kink}$ and a small positive $\gamma^{\prime}$ value. This scaling directly represents long-range part of critical behavior. The second one is given by
\begin{eqnarray}
S_{\chi}=\log\chi+\gamma=\frac{1}{6}\log n_{kink}+\gamma^{\prime}+\Delta S_{\chi}, \label{qc2}
\end{eqnarray}
for $\chi> n_{kink}$. This leads to
\begin{eqnarray}
S=\log L -2,\label{qc3}
\end{eqnarray}
for $\chi=L\rightarrow\infty$. The factor $\Delta S_{\chi}$ represents correction to short-range spin correlation. Although Eqs.~(\ref{qc}) and (\ref{qc2}) are only numerically confirmed in the Ising model, Eq.~(\ref{qc3}) is hold in the Ising and the $3$-state Potts model. Thus, their equations are expected to be fundamental ones irrespective of the value of the central charge.

Equation~(\ref{qc}) is quite analogous to the MPS scaling in Eq.~(\ref{mps}), but the definitions of $\chi$ are different with each other. In our case, $\chi$ is the truncation number of the SVD for the snapshot of 2D classical spin configuration, while $\chi$ represents the matrix dimension in the MPS. Although the MPS is also originated from the SVD, the role of $\chi$ on the scaling is quite different. In the MPS scaling, we need more and more matrix dimensions for large-$\xi$ cases. On the other hand, in the present case, it is enough to take $\chi=n_{kink}$ to catch the long-range correlation of the model. The relation between the length scale and $\chi$ are reversed in the MPS and our analysis.

Usually, 1D quantum sysytems can be mapped onto 2D classical ones by the Suzuki-Trotter decomposition. We consider that our observation potentially catches 1D quantum physics with the same universality class. Let us remember the transverse-field Ising chain that can be mapped onto the anisotropic 2D classical Ising model. Although we do not know the unique quantum correspondence of the isotropic Ising model, but we expect the presence of the quantum model. Then, multiplying a factor $(1/6)c{\cal A}$ into $S$ in Eq.~(\ref{qc3}) at $T_{c}$ leads to
\begin{eqnarray}
S^{classical}=\frac{1}{6}c{\cal A}S,
\end{eqnarray}
with $\xi=L$ and this may correspond to the entanglement entropy of 1D quantum systems in Eq.~(\ref{deformed}) (or Eq.~(\ref{CFT})).

Finally, we briefly touch on the presence of the topological term in our formalism~\cite{Kitaev,Levin}. Equation~(\ref{kink}) can be transformed into
\begin{eqnarray}
\left(\frac{1}{36}\log 2\right) n_{kink}=\frac{1}{12}\log L - \log\sqrt{2}. \label{topological}
\end{eqnarray}
This equation may be composed of the deformed CFT scaling and the topological term. Namely,
\begin{eqnarray}
n_{kink}\propto \frac{1}{6}c{\cal A}\log L -\log\sqrt{q}, \label{topo2}
\end{eqnarray}
where we assume ${\cal A}=1$. We start from the torus geometry due to the periodic boundary condition, and cut the trus in order to calculate the reduced density matrix. Then, the number of the cut for each direction is $1$. The relation~(\ref{topo2}) seems to be consistent with the numerical data for the $3$-state Potts model. When we take $c=4/5$ and $q=3$, Eq.~(\ref{topo2}) is given by $n_{kink}\propto 4\log_{3}L-15$. Then, we obtain $4\log_{3}L-15=1$, $2.64$, and $5$ for $L=81=3^{4}$, $128=2^{7}$, and $243=3^{5}$, respectively. On the other hand, the kink positions in Figs.~\ref{fig13}~(a) and (b) are located at $3$, $8\sim 9$, and $14\sim 15$. Thus, we find
\begin{eqnarray}
\frac{1}{3}n_{kink}\sim 4\log_{3}L-15.
\end{eqnarray}
Therefore, Eq.~(\ref{topo2}) is satisfied for two models with the different magnitudes of the central charge, respectively.

The second term in Eq.~(\ref{topo2}), $-\log\sqrt{q}$, would be a sign of topological nature in a 1D quantum system corresponding to the present 2D classical Ising model, although usually the topological entanglement entropy $S_{topo}$ is defined on the 2D quantum systems as $S=\alpha L+S_{topo}$. However, possible presence of topological effects even in 1D has been discussed recently in the Haldane phase of $S=1$ chains~\cite{Pollmann}. According to the definition of the entanglement gap, the gap separates low-energy topological levels from high-energy generic ones. Therefore, the doubly-degenerate edge modes are topologically protected, and this protection is coming from a set of symmetries. In the present case, on the other hand, the lowest-energy state below the entanglement gap represents ferromagnetic order. Then, the $Z_{q}$ symmetry is spontaneously broken. It would be important to clarify whether such a dual nature really exists or not, since the Haldane phase is not critical.

In summary, we have examined the entanglement entropy of the coarse-grained MC snapshots of 2D square-lattice Ising and $3$-state Potts models. Up to now, this type of entropy has not been examined yet, but we have found rich physical aspects hidden in the entropy. In particular near $T_{c}$, the entropy naturally produces the CFT scaling and the scaling for the entanglement support of the MPS approximation on a unified framework. In addition to the entropy, the entanglement spectrum also gives us the scaling relations due to the presence of the multiple entanglement gap. A key ingredient of these scaling relations is fractal nature of ferromagnetic islands near $T_{c}$ which have various length scale. Then, the SVD automatically decomposes the original snapshot into a set of images with different length scales, respectively. The kink structure of the entanglement gap characterizes this decomposition, and further suggests the possible presence of the topological term. Based on the present results, we are very interested in the detailed understanding of the duality between 1D quantum and 2D classical systems. Examinations of the scaling and the duality in much broader universality classes will be important future works.

\section*{Acknowledgements}

The author would like to thank K. Okunishi, N. Shibata and I. Maruyama for fruitful discussions and technical comments.


\begin{thebibliography}{99}

\bibitem{Bekenstein}
J. D. Bekenstein, Phys. Rev. D {\bf 7}, 2333 (1973).
\bibitem{Hawking}
S. W. Hawking, Phys. Rev. D {\bf 13}, 191 (1976).
\bibitem{Srednicki}
M. Srednicki, Phys. Rev. Lett. {\bf 71}, 666 (1993).
\bibitem{Holzhey}
C. Holzhey, F. Larsen, and F. Wilczek, Nucl. Phys. B {\bf 424}, 443 (1994).
\bibitem{Vidal}
G. Vidal, J. I. Latorre, E. Rico, and A. Kitaev, Phys. Rev. Lett. {\bf 90}, 227902 (2003).
\bibitem{Calabrese}
P. Calabrese and J. Cardy, J. Stat. Mech. {\bf 0406}, P002 (2004).
\bibitem{Plenio}
M. B. Plenio, J. Eisert, J. Drei{\ss}ig, and M. Cramer, Phys. Rev. Lett. {\bf 94}, 060503 (2005).
\bibitem{Wolf}
Michael M. Wolf, Phys. Rev. Lett. {\bf 96}, 010404 (2006).
\bibitem{Barthel}
T. Barthel, M. -C. Chung, and U. Schollw$\ddot{\rm o}$ck, Phys. Rev. A {\bf 74}, 022329 (2006).
\bibitem{Riera}
A. Riera and J. I. Latorre, Phys. Rev. A {\bf 74}, 052326 (2006).
\bibitem{Li}
Weifei Li, Leitan Ding, Rong Yu, Tommaso Roscilde, and Stephan Haas, Phys. Rev. B {\bf 74}, 073103 (2006).
\bibitem{Ostlund}
S. $\ddot{\rm O}$stlund and S. Rommer, Phys. Rev. Lett. {\bf 75}, 3537 (1995); S. Rommer and S. $\ddot{\rm O}$stlund, Phys. Rev. B {\bf 55}, 2164 (1997).
\bibitem{Tagliacozzo}
L. Tagliacozzo, Thiago. R. de Oliveira, S. Iblisdir, and J. I. Latorre, Phys. Rev. B {\bf 78}, 024410 (2008).
\bibitem{Huang}
Ching-Yu Huang and Feng-Li Lin, Phys. Rev. A {\bf 81}, 032304 (2010).
\bibitem{Calabrese2}
Pasquale Calabrese and Alexandre Lefevre, Phys. Rev. A {\bf 78}, 032329 (2008).
\bibitem{Thomale}
R. Thomale, A. Sterdyniak, N. Regnault, and B. Andrei Bernevig, Phys. Rev. Lett. {\bf 104}, 180502 (2010).
\bibitem{Kitaev}
A. Kitaev and J. Preskill, Phys. Rev. Lett. {\bf 96}, 110404 (2006).
\bibitem{Levin}
M. Levin and Xiao-Gang Wen, Phys. Rev. Lett. {\bf 96}, 110405 (2006).
\bibitem{Pollmann}
Frank Pollmann, Ari M. Turner, Erez Berg, and Masaki Oshikawa, Phys. Rev. B {\bf 81}, 064439 (2010).

\end{thebibliography}
\end{document}